\documentclass[prc,twocolumn,amsmath,amssymb,floatfix,nofootinbib]{revtex4}
\usepackage{mathrsfs,bm}
\usepackage{longtable,lscape}
\usepackage{txfonts}
\usepackage{amssymb}
\usepackage{indentfirst}
\usepackage{graphicx,,booktabs}
\usepackage{multirow}
\usepackage{color}
\usepackage{amssymb}
\definecolor{cover}{rgb}{0.77,0.87,0.88}
\definecolor{blueone}{rgb}{0.1,0.1,.7}
\definecolor{citec}{rgb}{0.14,0.47,0.09}
\definecolor{two}{rgb}{0.0,0.5,0.}
\definecolor{three}{rgb}{.5,.1,0.15}
\usepackage[bookmarks=true,bookmarksopen=false,plainpages=false,breaklinks=true,
   bookmarksnumbered=true,hypertexnames=false,
   filecolor=blue,urlcolor=three,menucolor=three,
   linkcolor=three,citecolor=blueone, colorlinks,
   anchorcolor=blue,runcolor=pink,frenchlinks=red
   pdfstartview=FitH,pdftitle=title,%
   pdfauthor=author]{hyperref}

\usepackage{relsize}
\usepackage{xspace}
\def\babar{\mbox{\slshape B\kern-0.1em{\smaller A}\kern-0.1em
    B\kern-0.1em{\smaller A\kern-0.2em R}}}

\allowdisplaybreaks[4]
\begin{document}
\title{Possible molecular states from the $N\Delta$ interaction }

\author{Zhi-Tao Lu, Han-Yu Jiang, Jun He}\email{Corresponding author: junhe@njnu.edu.cn}
\affiliation{Department of  Physics and Institute of Theoretical Physics, Nanjing Normal University, Nanjing 210097, People's Republic of China}

\date{\today}
\begin{abstract}
	
Recently, a hint  for dibaryon $N\Delta(D_{21})$  was observed at WASA-AT-COSY with a mass about $30\pm10$ MeV below the $N\Delta$ threshold. It has a relatively small binding energy compared with the $d^*(2380)$ and a width close to the width of the $\Delta$ baryon, which suggests that it may be a dibaryon in a molecular state picture. In this work, we study the possible $S$-wave  molecular states from the $N\Delta$ interaction within the quasipotential Bethe-Salpeter equation approach. The interaction is described by exchanging $\pi$, $\rho$, and $\omega$ mesons. With reasonable parameters, a $D_{21}$ bound state can be produced from the interaction. The results also suggest that there may exist two more  possible $D_{12}$ and $D_{22}$  states with smaller binding energies. The $\pi$ exchange is found to play the most important role to bind two baryons to form the molecular states. An experimental search for  possible $N\Delta(D_{12})$ and $N\Delta(D_{22})$ states will be helpful for understanding the hint of the dibaryon $N\Delta(D_{21})$.

\end{abstract}

\maketitle
\section{Introduction}

In the past two decades, the study of exotic  hadrons has become one of the most important topics in the community of  hadron physics.  The core issue of  hadron physics is to understand how quarks combine into a hadron. In the conventional quark model, a hadron is composed of $q\bar{q}$ as a meson or $qqq$ as a baryon. It is natural to expect the existence of  hadrons composed of more quarks, which are called exotic states. The deuteron can be also seen as a hadron, which is a quark system with six quarks, although we called it  a nucleus.  The existence of the nucleus and the hypernucleus inspires us to search for molecular states as  loosely bound states of  hadrons. Such a picture has been widely applied to interpret  the experimentally observed $XYZ$ particles and the hidden-charm pentaquarks $P_c$~\cite{Wang:2013cya,Guo:2017jvc,Aceti:2014uea,He:2013nwa,Wang:2014gwa,Sun:2011uh,Wu:2010jy,Yang:2011wz,Chen:2015loa,Roca:2015dva,He:2015cea,Karliner:2015ina,Liu:2019tjn,He:2019ify}.  More and more structures observed near thresholds of two hadrons give people more confidence about the existence of  molecular states. If we turn  back to the deuteron, which is a molecular state, it is interesting to study possible molecular states composed  of  two nucleons and/or its resonances, such as the systems $N\Delta$ and $\Delta\Delta$.

The hadron carrying baryon number $B=2$ is called dibaryon. The history of the study of dibaryons is even much longer than that of the $XYZ$ particles. Dyson and Xuong first predicted dibaryon states  in 1964 based on the SU(6) symmetry~\cite{Dyson:1964xwa} almost at the same time of the proposal of the quark model.   With a simple mass formula, the mass of deuteron was obtained as 1876 MeV, and the masses of dibaryon $\Delta\Delta(D_{03, 30})$  and  of  dibaryon $N\Delta(D_{12,21})$ were predicted at 2376 and 2176 MeV, respectively.  After observing an experimental hint in 1977~\cite{Kamae:1976as}, Kamae and Fujita made a calculation in the one-boson-exchange model at the hadronic level to reproduce an anomaly at 2380 MeV in the process $\gamma d\to p n$~\cite{Kamae:1976at}. The existence of the $\Delta\Delta(D_{03})$ was supported by many theoretical calculations especially the constituent quark model~\cite{Goldman:1989zj,Wang:1992wi,Yuan:1999pg,Li:2000cb}. The $N\Delta(D_{12})$ was also predicted in the literature~\cite{Mulders:1980vx,Mulders:1982da,Valcarce:2005em}.  The existence of the $N\Delta(D_{12})$ state was favored by  some early analyses of experimental data, such as the partial-wave analysis of the reaction $\pi^+ d\to pp$~\cite{Kravtsov:1984fw},    an analyses of $pp$ and $np$ scatterings by the SAID group~\cite{Arndt:1986jb},  and a study of the phase shifts  for the $N\Delta$ scattering with a nearby S-matrix pole based on the data of process $pp\to np\pi^+$~\cite{Hoshizaki:1992uh}.  However, the $N\Delta(D_{21})$ was not supported by the early calculation in the constituent quark model~\cite{Valcarce:2005em,Ping:2000cb,Ping:2008tp}. In Ref.~\cite{Green:1976sk}, the experimental data were also reproduced without the dibaryon.

After the efforts of more than  half a century on both the theoretical and  experimental sides\cite{Kamae:1976as,Kamae:1976at,Goldman:1989zj,Wang:1992wi,Yuan:1999pg,Li:2000cb,Sato:1982nr,Mulders:1980vx,Mulders:1982da,Valcarce:2005em,Kravtsov:1984fw,Arndt:1986jb,Hoshizaki:1992uh,Oka:1980ax,Ping:2000cb,Ping:2008tp,Maltman:1989qu,Ping:2000dx,Bashkanov:2008ih},   a candidate of dibaryon with $I(J^P)=0(3^+)$  carrying a mass of about 2370 MeV and a width of about 70 MeV was observed in the process $pp\rightarrow d\pi^0\pi^0$ at WASA-at-COSY~\cite{Adlarson:2011bh}, denoted as $d^*(2380)$. Later,  a series of measurements confirmed the existence of this state~\cite{Adlarson:2012fe,Adlarson:2013usl,Adlarson:2014pxj}. Such  a state was also confirmed by a recent measurement  within the Crystal Ball at MAMI, where the photoproduction process was performed~\cite{Bashkanov:2019mpj}. The observation of the $d^*(2380)$ attracts much attention from theorists, and a large number of  interpretations were proposed to understand its properties and internal structure~\cite{Gal:2014zia,Gal:2013dca,Huang:2013nba,Huang:2014kja,Haidenbauer:2011za,Park:2015nha,Gal:2016bhp,Dong:2015cxa,Bashkanov:2015xsa,Chen:2014vha,Dong:2016rva,Dong:2017geu,Huang:2015nja}. Because the $\Delta$ signal can be found in the final states of its decay, one may guess that it is a $\Delta\Delta$ bound state. However, such an  assumption leads to a binding energy of about 80 MeV considering that the mass of thte $\Delta$ baryon is about 1232 MeV. Such a large binding energy prefers a compact hexaquark instead of a bound state of two $\Delta$ baryons. The conclusion is further supported by  the relatively smaller width of 70 MeV of  $d^*(2380)$, which is even smaller than the width of one $\Delta$ baryon, about 120 MeV.  With a Faddeev equation calculation, Gal and Garcilazo proposed that  $d^*(2380)$ is from a three-body $N\Delta\pi$ system ~\cite{Gal:2014zia,Gal:2013dca}. In their study, the  $N\Delta(D_{12, 21})$ was also studied in the three-body $NN\pi$ interaction and found slightly below the $N\Delta$ threshold.

Recently, an isotensor dibaryon $N\Delta$ with quantum numbers $IJ^{P}=21^+ (D_{21})$ with a mass of $2140(10)$ MeV and a width of $110(10)$ MeV was reported at WASA-at-COSY~\cite{Adlarson:2018bbv}.   Its mass is about 30$\pm10$ MeV below the $N\Delta$ threshold. Considering that the width of nucleon is zero (for protons) or very small (for neutrons) and the $\Delta$ baryon has a width of about 120 MeV, the width of this $N\Delta(D_{21})$ state is almost the sum  of nucleon and $\Delta$ baryon. Hence, compared with the $d^*(2380)$, such a state is obviously  consistent with the molecular state picture.
In Ref.~\cite{Huang:2018izl}, the authors studied the $N\Delta$ states in the constituent quark model, and found that it is less likely for $N\Delta(D_{21})$ than $N\Delta(D_{12})$ to form a bound state.
In this work, with the help of the effective Lagrangians, we will construct  the  interaction in the one-boson-exchange model, and insert it into the quasipotential Bethe-Salpeter equation (qBSE) to find $S$-wave bound states from the $N\Delta$ interaction.

The paper is organized as follows. After the introduction, we present the effective Lagrangians and  relevant coupling constants to describe the $N\Delta$ interaction, with which we deduce the potential. And the qBSE is also briefly introduced. In Sec.~\ref{Sec: results}, we will present the numerical results, and the contributions from different exchanges and diagrams are also discussed. Finally, the article ends with a summary in Sec.~\ref{Sec: summary}.

\section{Theoretical frame}\label{Sec: Formalism}

\begin{figure}[h!]
\centering
\includegraphics[bb=75 680 600 770, clip,scale=0.53]{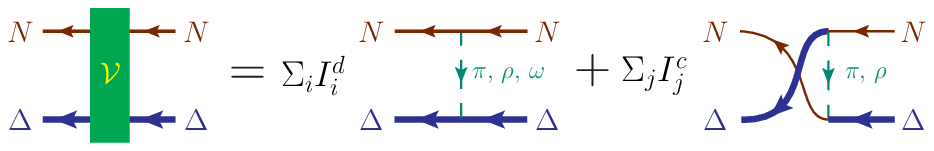}
\caption{The diagrams for the direct (left) and cross (right) potentials. The thin (brown) and thick (blue) lines are for $N$ and $\Delta$ mesons, respectively. $I^{(d)}_i$ and $I^{(c)}_i$ are the flavor factors for the direct and cross diagram, respectively, with $i$ exchange, which is explained in the text.}
\label{V}
\end{figure}

In the current work, we will describe the  $N \Delta$ interaction in the one-boson-exchange model, in which the interaction is usually mediated by the exchange of the light mesons including   pseudoscalar mesons ($\pi$ and $\eta$), vector mesons ($\rho$, $\omega$, and $\phi$), and  scalar meson $\sigma$. The coupling of the $\eta$ meson and nucleon is small~\cite{Li:1998ni,Tiator:1994et,Kirchbach:1996kw,Zhu:2000eh,He:2010ii,He:2008uf,He:2008ty,Zhong:2007fx}, and the coupling of the $\phi$ meson and nucleon is suppressed according to the OZI rule. Besides, we do not consider the scalar meson exchange as done in Refs.~\cite{Matsuyama:2006rp,Ronchen:2012eg}. Hence, we only consider the exchanges of $\pi$, $\rho$, and $\omega$ mesons in the calculation. There are two diagrams for the $N\Delta$ interaction as shown in Fig.~\ref{V}. In the cross diagram, the $\omega$ exchange is forbidden due to  conservation of  isospin.

We need the Lagrangians for the vertices of nucleon, $\Delta$ baryon, and pseudoscalar meson $\pi$, which are written as~\cite{Matsuyama:2006rp,Ronchen:2012eg}
\begin{eqnarray}
\mathcal{L}_{NN\pi}&=& -\frac{g_{NN\pi}}{m_{\pi}}~\bar{N}\gamma^5\gamma^{\mu}{\bm\tau}\cdot\partial_{\mu}{\bm\pi}N,\\
\mathcal{L}_{\Delta\Delta\pi}&=& \frac{g_{\Delta\Delta\pi}}{m_{\pi}}~\bar{\Delta}_{\mu}\gamma^5\gamma^{\nu}{\bm T}\cdot\partial_{\nu}{\bm \pi}\Delta^{\mu},\\
\mathcal{L}_{N\Delta\pi}&=& \frac{g_{N\Delta\pi}}{m_{\pi}}~\bar{\Delta}^{\mu}{\bm S}^{\dag}\cdot\partial_{\mu}{\bm \pi}N+{\rm H.c.},
\end{eqnarray}
where the $N$, $\Delta$, and $\pi$ are nucleon, $\Delta$ baryon, and pion meson fields. The coupling constants $g^2_{NN\pi}/4\pi=0.08$, $g_{\Delta\Delta\pi}=1.78$, and $g_{N\Delta\pi}=-2.049$, which were obtained from fitting the experimental data in Refs.~\cite{Sato:1996gk,Matsuyama:2006rp,Ronchen:2012eg}.

The Lagrangians for the vertices of nucleon, $\Delta$ baryon, and vector meson $\rho/\omega$ are written as~\cite{Matsuyama:2006rp,Ronchen:2012eg},
\begin{eqnarray}
\mathcal{L}_{NN\rho}&=& -g_{NN\rho}~\bar{N}[\gamma^{\mu}-\frac{\kappa_{\rho}}{2m_{N}}\sigma^{\mu\nu}\partial_{\nu}]{\bm \tau}\cdot{\bm \rho}_{\mu}N,\nonumber\\
\mathcal{L}_{NN\omega}&=& -g_{NN\omega}~\bar{N}[\gamma^{\mu}-\frac{\kappa_{\omega}}{2m_{N}}\sigma^{\mu\nu}\partial_{\nu}]{\omega}_{\mu}N,\nonumber\\
\mathcal{L}_{\Delta\Delta\rho}&=& -g_{\Delta\Delta\rho}~\bar{\Delta}_{\tau}(\gamma^{\mu}-\frac{\kappa_{\Delta\Delta\rho}}{2m_{\Delta}}\sigma^{\mu\nu}\partial_{\nu}){\bm \rho}_{\mu}\cdot {\bm T}\Delta^{\tau},\nonumber\\
\mathcal{L}_{\Delta\Delta\omega}&=& -g_{\Delta\Delta\omega}~\bar{\Delta}_{\tau}(\gamma^{\mu}-\frac{\kappa_{\Delta\Delta\omega}}{2m_{\Delta}}\sigma^{\mu\nu}\partial_{\nu})\omega^\mu\Delta^{\tau},\nonumber\\
\mathcal{L}_{N\Delta\rho}&=& -i\frac{g_{N\Delta\rho}}{m_{\rho}}~\bar{\Delta}^{\mu}\gamma^5\gamma^{\nu}{\bm S}^{\dag}\cdot{\bm \rho}_{\mu\nu}N+{\rm H.c.},
\end{eqnarray}
where  ${\bm \rho}_{\mu\nu}=\partial_{\mu}{\bm \rho}_{\nu}-\partial_{\nu}{\bm \rho}_{\mu}$, and $\rho$ or $\omega$ denotes the $\rho$ or $\omega$ meson field. The coupling constants  are $g_{NN\rho}=-3.1$, $g_{\Delta\Delta\rho}=4.9$, $g_{N\Delta\rho}=6.08$,  $\kappa_{\rho}=1.825$, $\kappa_{\omega}=0$, $\kappa_{\Delta\Delta\rho}=6.1$, cited from Refs.~\cite{Sato:1996gk,Ronchen:2012eg,Matsuyama:2006rp}. The coupling constants for the $\omega$ meson can be related to these for the $\rho$ meson with  SU(3) symmetry as $g_{NN\omega}=3g_{NN\rho}$, $g_{\Delta\Delta\omega}=3/2g_{\Delta\Delta\rho}$, and $\kappa_{\Delta\Delta\omega}=\kappa_{\Delta\Delta\rho}$.
In addition, the $T$ and the $S$ matrices are provided as follows,
\begin{eqnarray}
{\bm T}\cdot{\bm\varphi}&=&\sqrt{\frac{4}{15}}\left(\begin{array}{cccc}
\frac{3}{2}\varphi^0 & \sqrt{\frac{3}{2}}\varphi^+& 0& 0\\
\sqrt{\frac{3}{2}}\varphi^- & \frac{1}{2}\varphi^0 & \sqrt{2}\varphi^+& 0\\
0 & \sqrt{2}\varphi^- & -\frac{1}{2}\varphi^0 &\sqrt{\frac{3}{2}}\varphi^+
\\ 0 & 0 &\sqrt{\frac{3}{2}}\varphi^- &-\frac{3}{2}\varphi^0
\end{array}\right),\\
{\bm S}\cdot{\bm\varphi}&=&\left(\begin{array}{cccc}
-\varphi^-& \sqrt{\frac{2}{3}}\varphi^0&\sqrt{\frac{1}{3}}\varphi^+ &0\\
0&-\sqrt{\frac{1}{3}}\varphi^-&\sqrt{\frac{2}{3}}\varphi^0&\varphi^+
\end{array}\right),
\end{eqnarray}
where $\phi=\pi$ or $\rho$, and the $+$, $-$, and $0$ denote the charges of the mesons.

Using the Lagrangians above, the potential of the $N\Delta$ interaction can be constructed as,
\begin{align}
i{\cal V}^d_\pi&=I^d_\pi\frac{g_{NN\pi}g_{\Delta\Delta\pi}}{m_{\pi}^2}~\bar{u}(k_1')\gamma^5\gamma^{\mu}q_{\mu}u(k_1)\nonumber\\
&\cdot\bar{u}^{\alpha}(k'_2)\gamma^5\gamma^{\nu}q_{\nu}u_{\alpha}(k_2)~iP_\pi(q^2),\label{V1}\\
i{\cal V}^d_\rho&=-I^d_\rho g_{NN\rho}g_{\Delta\Delta\rho}~\bar{u}(k'_1)\left(\gamma_{\mu}-\frac{\kappa_{\rho}}{2m_{N}}i\sigma_{\mu\alpha}q^{\alpha}\right)u(k_1)\nonumber\\
&\cdot\bar{u}^{\kappa}(k_2')\left(\gamma_{\mu}+\frac{\kappa_{\Delta\Delta\rho}}{2m_{\Delta}}i\sigma_{\nu\beta}q^{\beta}\right)u_{\kappa}(k_2)~iP^{\mu\nu}_\rho(q^2),\\
i{\cal V}^d_\omega&=-I^d_\omega g_{NN\omega}g_{\Delta\Delta\omega}~\bar{u}(k'_1)\left(\gamma_{\mu}-\frac{\kappa_{\omega}}{2m_{N}}i\sigma_{\mu\alpha}q^{\alpha}\right)u(k_1)\nonumber\\
&\cdot\bar{u}^{\kappa}(k_2')\left(\gamma_{\mu}+\frac{\kappa_{\Delta\Delta\omega}}{2m_{\Delta}}i\sigma_{\nu\beta}q^{\beta}\right)u_{\kappa}(k_2)~iP^{\mu\nu}_\omega(q^2),\\
i{\cal V}^c_\pi&=I^c_\pi\frac{-g^2_{N\Delta\pi}}{m^2_{\pi}}~\bar{u}^{\mu}(k'_2)q_{\mu}u(k_1) \bar{u}(k'_1)q_{\nu}u^{\nu}(k_2)~iP_\pi(q^2),\\
i{\cal V}^c_\rho&=-I^c_\rho\frac{g^2_{N\Delta\rho}}{m^2_{\rho}}~\bar{u}_{\alpha}(k_2')\gamma^5\left(\gamma^{\mu}q^\alpha-g^{\mu\alpha}\rlap\slash q\right)u(k_1)\nonumber\\
&\cdot \bar{u}(k'_1)\left(\gamma^{\nu}q^\beta-g^{\nu\beta}\rlap\slash q\right)\gamma^5u^{\beta}(k_2)~iP^{\mu\nu},\label{V5}
\end{align}
where the $u$ and $u^\alpha$ are the spinor for nucleon and the Rarita-Schwinger vector-spinor for $\Delta$ baryon, respectively, and $q$, $k_{(1,2)}$, and $k'_{(1,2)}$ are the momenta of the exchange meson,  and the initial and final nucleons or $\Delta$ baryons.
The flavor factors $I^{(d,c)}_i$ for certain meson exchange and total isospin are presented in Table~\ref{flavor factor}.
\renewcommand\tabcolsep{0.345cm}
\renewcommand{\arraystretch}{1.5}
\begin{table}[h!]
\caption{The flavor factors $I^{(d,c)}_i$ for certain meson exchange  and total isospin.  \label{flavor factor}}
\begin{tabular}{c|ccccc}\bottomrule[2pt]
& $I^d_\pi$&$I^d_\rho $ & $I^d_\omega $ & $I^c_\pi $ & $I^c_\rho $\\\hline
$I=1$&$-\sqrt{15}/3$ &$-\sqrt{15}/3$ &$1$&$-1/3$&$-1/3$\\
 $I=2$&$\sqrt{15}/5$ &$\sqrt{15}/5$ &$1$&$1$&$1$\\
\toprule[2pt]
\end{tabular}

\end{table}

The propagators of  exchanged mesons are of the usual forms of
$
P_e(q^2)={if_{i}(q^{2})}/{(q^2-m_e^2)}$ 
and 
$P_e^{\mu\nu}(q^2)={if_{i}(q^{2})(-g^{\mu\nu}+{q^{\mu}q^{\nu}}/{m_e^2})}/{(q^2-m_e^2)}
$
where $m_e$ is the mass of the exchanged meson. The form factor $f_{i}(q^{2})$ is used to compensate  the off-shell effect of the exchanged meson. In this work, we introduce four types of form factors to check the effect of the
form factor on the results. These have the forms~\cite{He:2019rva},
\begin{align}
f_1(q^2)&=\frac{\Lambda_e^2-m_e^2}{\Lambda_e^2-q^2},\label{FF1}\\
f_2(q^2)&=\frac{\Lambda_e^4}{(m_e^2-q^2)^2+\Lambda_e^4},\label{FF2}\\
f_3(q^2)&=e^{-(m_e^2-q^2)^2/\Lambda_e^4},\\
f_4(q^2)&=\frac{\Lambda_e^4+(q^2_t-m_e^2)^2/4}{[q^2-(q^2_t+m_e^2)/2]^2+\Lambda_e^4},\label{FF4}
\end{align}
where the  $q_t^2$ denotes the value of $q^2$ at the kinematical threshold. The cutoff is parametrized as a form of $\Lambda_e=m+\alpha_e~0.22$ GeV. In the current work, we  change the $q^2$ in the cross diagram to $-|q^2|$ to avoid the singularities as done in Ref.~\cite{Gross:2008ps}.

In this work, we  adopt the Bethe-Salpeter equation to obtain the $N\Delta$  scattering amplitudes. With the spectator quasipotential approximation~\cite{Gross:2010qm,He:2012zd,He:2011ed}, the Bethe-Salpeter equation was reduced into a three-dimensional qBSE, which is further reduced into a one-dimensional  equation with fixed spin-parity $J^P$ after a partial-wave decomposition as~\cite{He:2015cca,He:2015mja,He:2017aps},
\begin{eqnarray}
i{\cal M}^{J^P}_{\lambda'\lambda}({\rm p}',{\rm p})
&=&i{\cal V}^{J^P}_{\lambda',\lambda}({\rm p}',{\rm
p})+\sum_{\lambda''}\int\frac{{\rm
p}''^2d{\rm p}''}{(2\pi)^3}\nonumber\\
&\cdot&
i{\cal V}^{J^P}_{\lambda'\lambda''}({\rm p}',{\rm p}'')
G_0({\rm p}'')i{\cal M}^{J^P}_{\lambda''\lambda}({\rm p}'',{\rm
p}),\quad\quad \label{Eq: BS_PWA}
\end{eqnarray}
where the sum extends only over nonnegative helicity $\lambda''$.   With the spectator approximation, the $G_0({\rm p}'')$ is reduced from the 4-dimensional  propagator $G^{4D}_0({ p}'')$, which can be written down in the center-of-mass frame with $P=(W,{\bm 0})$ as,
\begin{align}
	G^{4D}_0({ p}'')&=\frac{\delta^+(p''^{~2}_\Delta-m_\Delta^{2})}{p''^{~2}_N-m_N^{2}}\nonumber\\
&=\frac{\delta^+(p''^{0}_\Delta-E_\Delta({\rm p}''))}{2E_\Delta({\rm p''})[(W-E_\Delta({\rm
p}''))^2-E_N^{2}({\rm p}'')]}.\label{G0}
\end{align}
Obviously,  the $\delta$ function will reduce the four-dimensional integral equation to a three-dimensional one, and  Eq.~(\ref{Eq: BS_PWA}) can be obtained after partial-wave decomposition.  Here, as required by the spectator approximation, the heavier $\Delta$ baryon  is on shell, which satisfies  $p''^0_\Delta=E_{\Delta}({\rm p}'')=\sqrt{
m_{\Delta}^{~2}+\rm p''^2}$ as suggested  by the $\delta$ function in Eq.~(\ref{G0}). The $p''^0_N$ for the lighter nucleon is then $W-E_{\Delta}({\rm p}'')$. Here and hereafter, the definition ${\rm p}=|{\bm p}|$ is adopted.

The partial-wave potential is defined with the potential of the interaction obtained  above as
\begin{eqnarray}
{\cal V}_{\lambda'\lambda}^{J^P}({\rm p}',{\rm p})
&=&2\pi\int d\cos\theta
~[d^{J}_{\lambda\lambda'}(\theta)
{\cal V}_{\lambda'\lambda}({\bm p}',{\bm p})\nonumber\\
&+&\eta d^{J}_{-\lambda\lambda'}(\theta)
{\cal V}_{\lambda'-\lambda}({\bm p}',{\bm p})],
\end{eqnarray}
where $\eta=PP_1P_2(-1)^{J-J_1-J_2}$ with $P$ and $J$ being parity and spin for system, nucleon  or $\Delta$ baryon. The initial and final relative momenta are chosen as ${\bm p}=(0,0,{\rm p})$  and ${\bm p}'=({\rm p}'\sin\theta,0,{\rm p}'\cos\theta)$. The $d^J_{\lambda\lambda'}(\theta)$ is the Wigner $d$-matrix. 

In our  qBSE approach, the $\Delta$ baryon is set on-shell while the nucleon can still  be off-shell.  Hence, we introduce a form factor into the propagator to reflect the off-shell effect as an exponential regularization,
$
G_{0}(p)\rightarrow G_{0}(p)[e^{-(k^{2}_{1}-m^{2}_{1})^{2}/\Lambda^{4}_{r}}]^{2},
$
where the $k_{1}$ and  $m_{1}$ are the momentum and the mass of the nucleon. With such regularization, the integral equation is convergent even if we do not consider the form factor into the propagator of  the exchanged meson.  The cutoff $\Lambda_{r}$ is parametrized as in the $\Lambda_{e}$ case, that is,  $\Lambda_{r}=m_{e}+\alpha_{r}~0.22$~GeV with $m_{e}$ being the mass of  the exchanged meson and $\alpha_{r}$ serving the same function as the parameter $\alpha_{e}$.  The $\alpha_e$ and $\alpha_r$ play  analogous roles in the calculation of the binding energy. Hence, we take these two parameters as a parameter $\alpha$ for simplification.

\section{Numerical results}\label{Sec: results}

\begin{figure}[h!]
  \centering
 \includegraphics[bb=12 50 600 880, clip,scale=0.425]{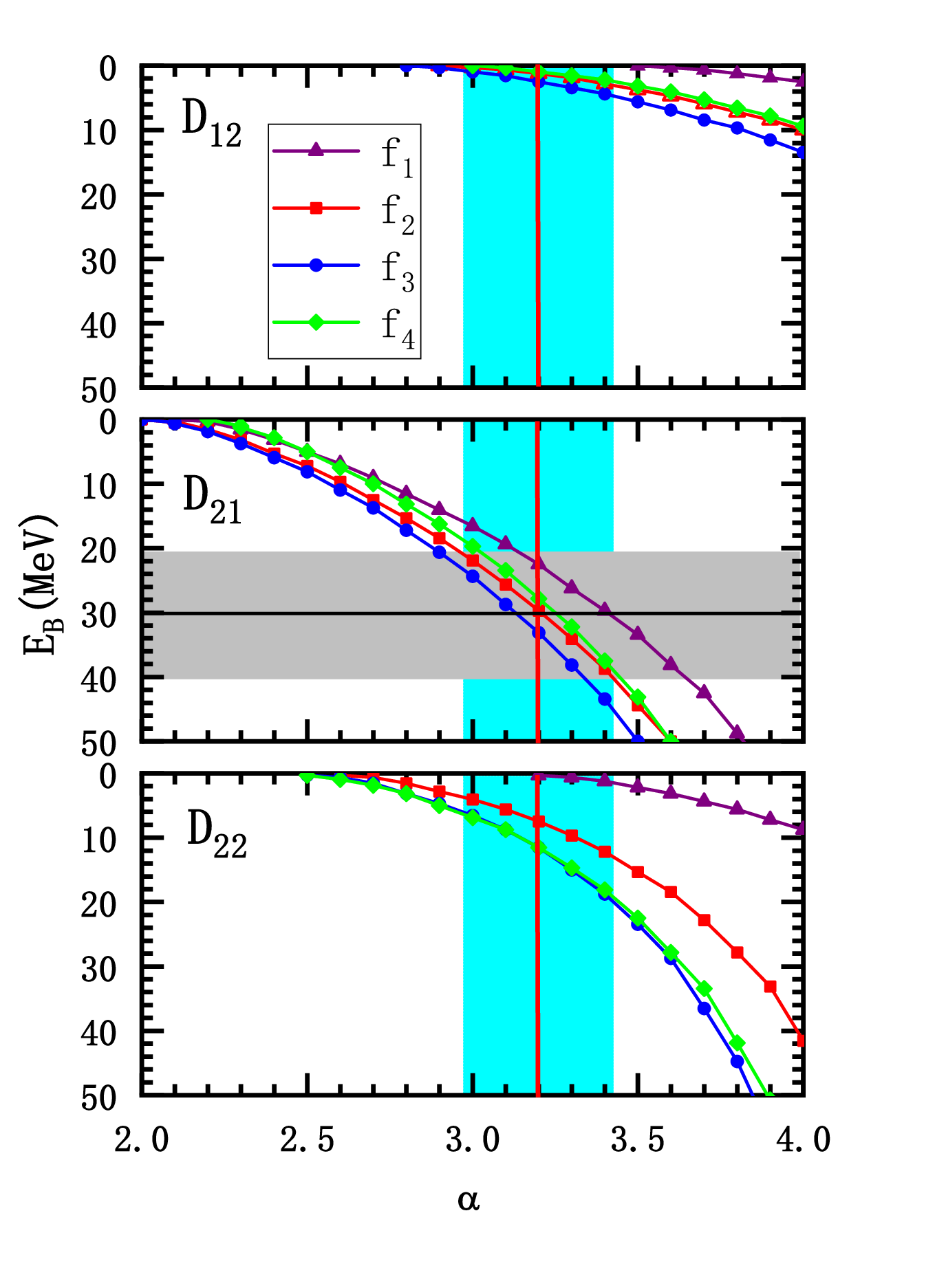}
\put(-203,255){\color{three}\large (a)}\put(-203,150){\color{three}\large(b)}\put(-203,45){\color{three}\large(c)}
  \caption{The variation of the binding energy $E_{B}=M_{th}-W$  on parameter $\alpha$ with $M_{th}$ and $W$ being the $N\Delta$ threshold and  position of  bound states. The triangle (purple), square (red), circle (blue), and diamond (green) and the corresponding lines are for  form factors of the types in Eqs.~(\ref{FF1}-\ref{FF4}). The horizontal line and the gray band in middle panel are for the experimental mass and its uncertainties observed at WASA-at-COSY~\cite{Adlarson:2018bbv}. The red line and cyan band are for the $\alpha$  determined by the experiment with form factor $f_2$.    }\label{total}
 \end{figure}

The scattering amplitude of the $N\Delta$ interaction can be obtained
by inserting the potential kernel in Eqs.~(\ref{V1}-\ref{V5})
into the qBSE in Eq.~(\ref{Eq: BS_PWA}). The bound state can be
searched as the pole in the real axis of the complex energy plane below the threshold. In the
current work, we will consider four $S$-wave states from the $N\Delta$
interaction, $D_{11}$, $D_{12}$, $D_{21}$, and $D_{22}$, with isospin spin $IJ=11$,
12, 21, and 22, respectively.  The results with the variation of the
parameter $\alpha$ are presented in Fig.~\ref{total}.

Among the four $S$-wave states considered, three bound states are produced from the $N\Delta$ interaction, that is, $D_{12}$, $D_{21}$, and $D_{22}$.  The $D_{21}$ state, which hint was observed at WASA-at-COSY, appears at an $\alpha$ of about two, and its binding energy increases with the increase of  $\alpha$. The experimental value of  the binding energy can be reached at $\alpha$ of about 3 to 3.5. In the figure, we present the experimental results of the mass and corresponding  uncertainty as a horizontal line and a gray band in the middle panel for reference. The values of an $\alpha$ can be determined by  comparing  the theoretical result and experiment. Here, we take the results with $f_2$ as an example, the determined  value of $\alpha$ and its uncertainty are shown as a red vertical line and a cyan band.  For $f_{2,3,4}$, two other bound states appears at larger $\alpha$, 2.5 and 2.8, for the $D_{12}$ and $D_{22}$ states, respectively. For $f_1$,  values of $\alpha$  about 0.5 larger are needed to produce these two states.  If we choose the value of $\alpha$ for $f_{2}$ as shown in figure as a red line,  the binding energies of the $D_{12}$ and $D_{22}$ states are about 2 and 8 MeV, respectively.  After considering the uncertainties, the binding energies of these two states are  several and  ten MeV, respectively. Hence,  the $D_{12}$ and $D_{22}$ states  are bound much more shallowly than the $D_{21}$ state. In the current work, we consider four types of the form factors as shown in the figure. As suggested by the results, the different choices of the form factor do not affect the conclusion obtained above with $f_2$.

\begin{figure}[h!]
  \centering
  \includegraphics[bb=50 30 710 490, clip,scale=0.39]{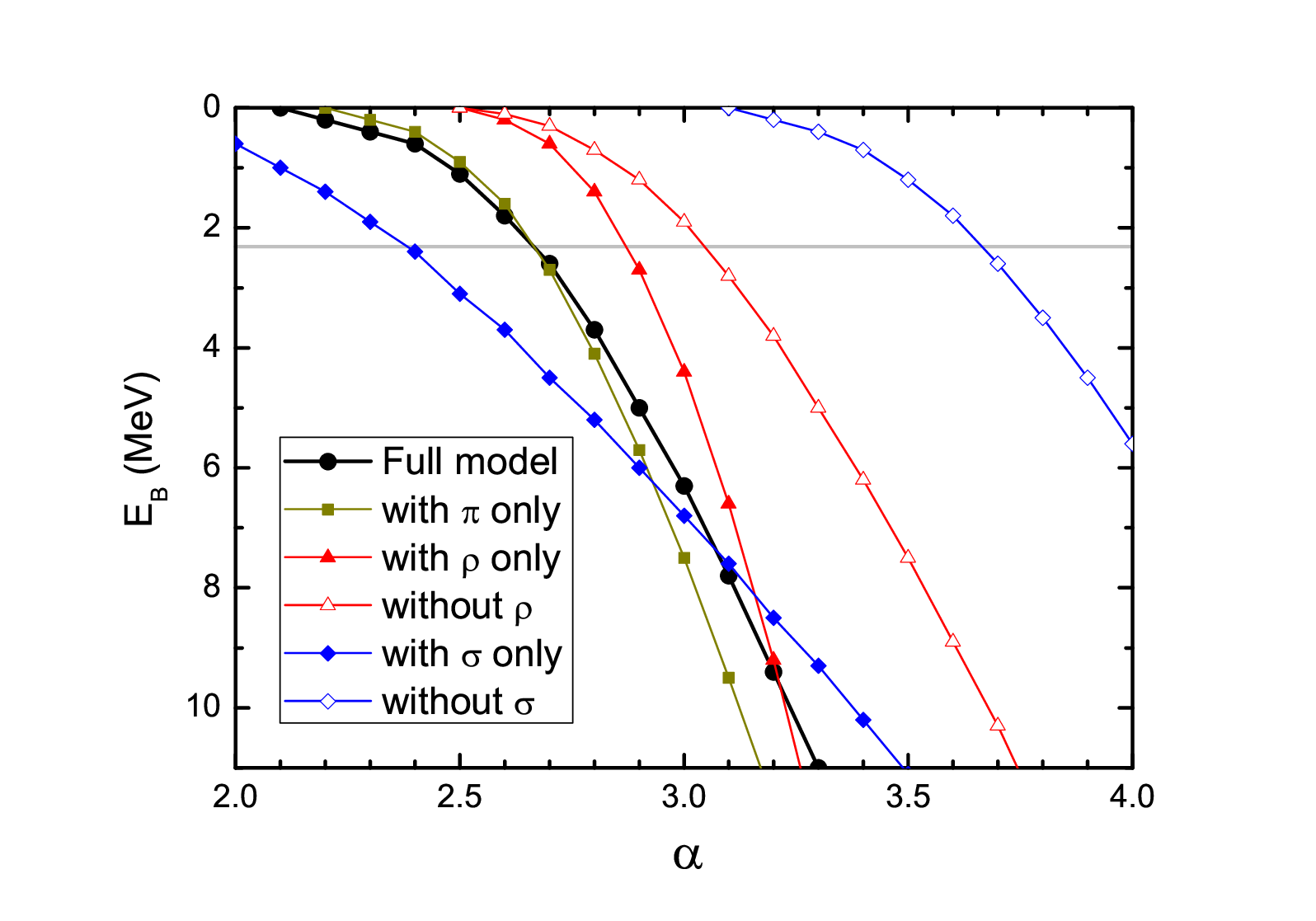}
  \caption{The variation of the binding energy $E_{B}$  for the $NN$ interaction with isospin $I=0$ and spin $S=1$ with $f_2(q^2)$ in Eq.~(\ref{FF2}). The horizontal line is for the experimental mass of the deuteron. 
 }\label{NN}
 \end{figure}

In the former discussions we presented the results for the $N\Delta$ interaction. The $NN$ scattering has been studied explicitly in Refs.~\cite{Gross:2008ps,Gross:2010qm} by Gross and his collaborators with the same spectator approximation adopted in the current work. Because  there are some differences in the explicit treatment between the current work and Refs.~\cite{Gross:2008ps,Gross:2010qm}, it is interesting to see if the deuteron can be reproduced with current Lagrangians and theoretical frames.  The potential can be obtained easily by replacing $\Delta$ by $N$, and the $\sigma$ exchange is introduced by a Lagrangian ${\cal L}_{\sigma NN}=g_{\sigma NN} \bar{N}N\sigma$ with a coupling constant  $g_{\sigma NN}\approx 5$~\cite{Oset:2000gn,Machleidt:1987hj}. In Fig.~\ref{NN}, we present the results for the $NN$ interaction with isospin $I=0$ and spin $J=1$ with $f_2(q^2)$.  It is found that,  with an $\alpha$ of about 2 the bound state was produced from the $NN(D_{01})$ interaction, which can be related to the deuteron.  Considering that the $NN$ and $N\Delta$ interactions are different, one can say that the $\alpha$ of about 3.2 adopted in the $N\Delta$ interaction is consistent with the $\alpha$ value used to reproduce the deuteron, about 2.7.  With only one of  the $\pi$, $\rho$, and $\sigma$ exchanges, the bound state can be found in the range of the parameter considered here, while with only the $\omega$ exchange no bound state  can be produced. It suggests that the $\pi$, $\rho$, and $\sigma$ exchanges provide attractive force.   If we remove the contribution from the $\omega$ exchange, a bound state will appear below $\alpha=2$, which concludes that  the $\omega$ exchange provides a repulsive force. Without the $\pi$ exchange, the bound state will disappear, while if we remove the  $\rho$ or $\sigma$ exchange, the bound state  still remains. It suggests that the $\pi$ exchange is essential to cancel the repulsive $\omega$ exchange.  Moreover, the result with the $\pi$ exchange only is very close to that with the full model.  Hence, the $\pi$ exchange is crucial to reproduce the deuteron. Such results are consistent with the usual conclusion of the OBE model of the nuclear force~\cite{Machleidt:1987hj}. We would like to remind the reader that,  compared with the works by Gross $et\ al.$~\cite{Gross:2008ps,Gross:2010qm} the calculation here is very crude  and we do not fit experimental data of $NN$ scattering either. It is  given only to show that our approach can give the basic results of  the nuclear force.  

In Fig.~\ref{total}, we choose a value of the parameter $\alpha=3.2$, which is  determined from the experimental mass. If we choose the parameter $\alpha=2.7$, which is required to  reproduce the deuteron, the conclusion for the $N\Delta$ interaction will change a little. The binding energy of the $D_{21}$ state  reduces to about 10~MeV, and the $D_{22}$ state has a very small binding energy, about 1 MeV. With such a parameter, the $D_{12}$ may disappear.

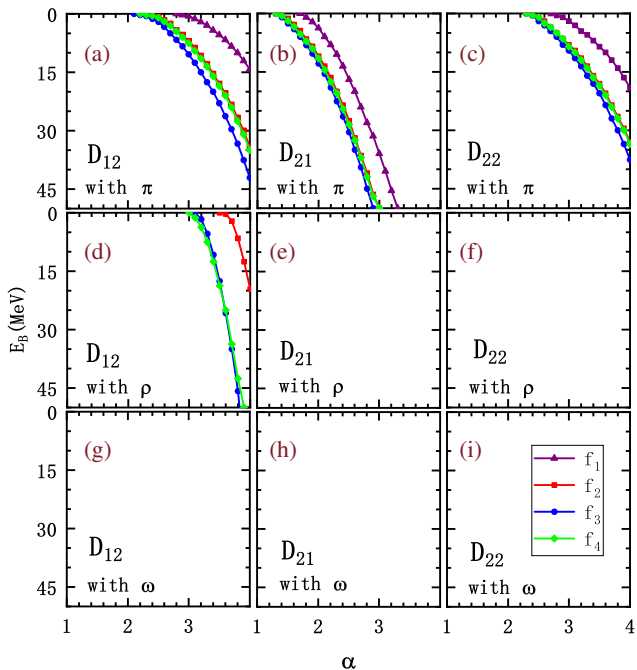
\begin{figure}[ht!]
  \includegraphics[bb=20 20 500 435,clip,scale=0.6]{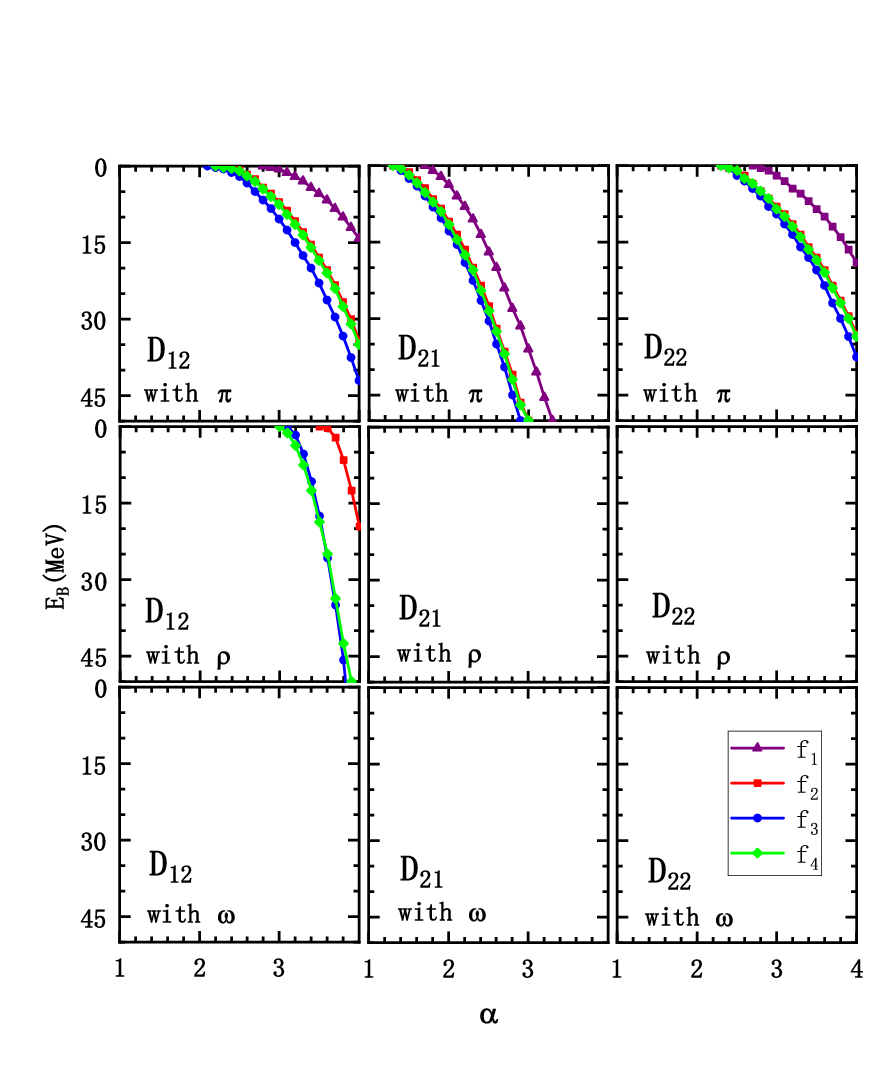}
  \put(-260,230){\color{three} (a)}\put(-190,230){\color{three}(b)}\put(-118,230){\color{three}(c)}
    \put(-260,155){\color{three} (d)}\put(-190,155){\color{three}(e)}\put(-118,155){\color{three}(f)}
        \put(-260,80){\color{three} (g)}\put(-190,80){\color{three}(h)}\put(-118,80){\color{three}(i)}
  \caption{The binding energy $E_{B}$ with variation of the $\alpha$ with exchange of only one meson. }\label{exchanges}
 \end{figure}

In the current work, we consider  three exchanges of the $\pi$, $\rho$, and $\omega$ mesons. In Fig.~\ref{exchanges} we present the results with only one exchange to discuss the  role played by each exchange.
Here we only present the results for three states which are bound by the interaction. For the $D_{12}$ state, the bound state can not be produced only with the $\omega$ exchange. With the $\rho$ exchange, the bound state still exists but appears at larger $\alpha$ of about 3.0. It suggests that the attraction is very weak compared with the full model, and a larger value of  $\alpha$ is needed to compensate  it.  For the results with only the $\pi$ exchange, one can find that the binding even becomes stronger than that with all three exchanges.  The explicit analysis suggests that the $\omega$ exchange will weaken the attraction, which leads to the larger $\alpha$ needed in the full model, which is also analogous to the deuteron case. Hence, for the $D_{12}$ state, the main attraction is from the $\pi$ exchange as in the deuteron case. The $\rho$ exchange  provides marginal attraction while inclusion of the  $\omega$ exchange  weakens the attraction. For the $D_{21}$ and $D_{22}$ states, only with $\pi$ exchange, the bound states  can be  produced, but the $\alpha$ needed is smaller than for the full model.
It suggests that the $\pi$ exchange plays the most important role in producing the bound states as for the $D_{12}$ state.

\begin{figure}[htp]
  \includegraphics[bb=15 30 390 320,clip,scale=0.64]{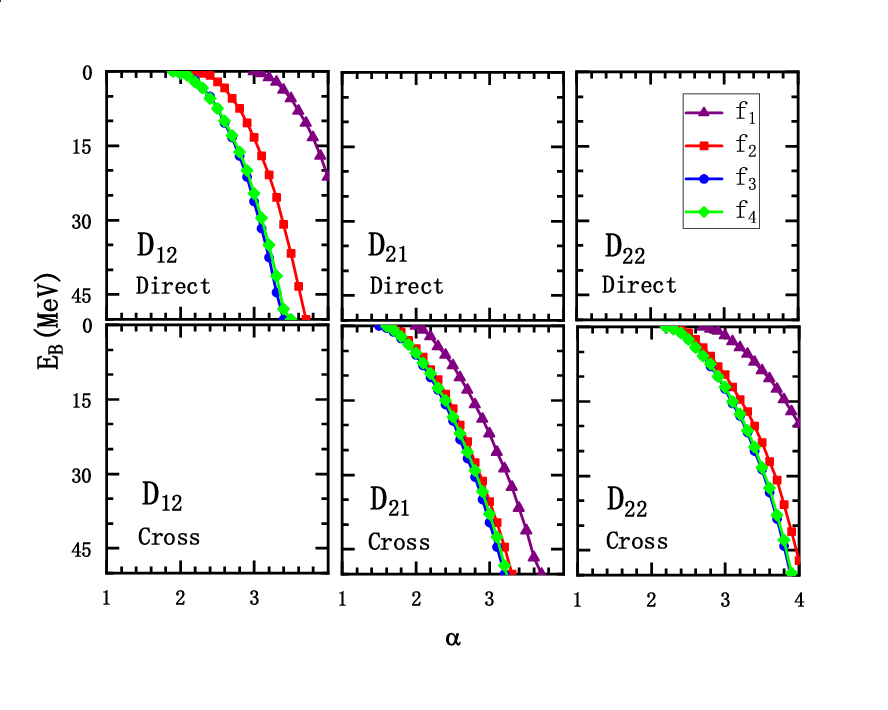}
    \put(-210,165){\color{three} (a)}\put(-137,165){\color{three}(b)}\put(-65,165){\color{three}(c)}
    \put(-210,85){\color{three} (d)}\put(-137,85){\color{three}(e)}\put(-65,85){\color{three}(f)}
  \caption{The binding energy $E_{B}$ with variation of the $\alpha$ for direct and cross diagrams.}\label{graph}
 \end{figure}
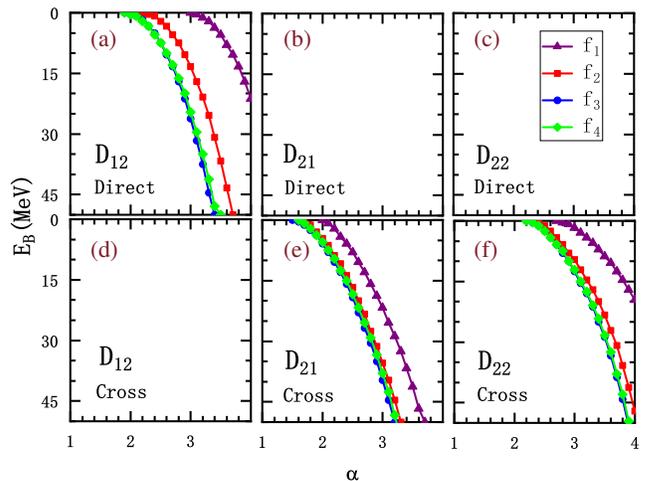

In our model, two diagrams are considered for the interaction, that is, the direct and cross diagrams as shown in Fig.~\ref{V}. In Fig.~\ref{graph}, we present the results with only one  diagram. For different states different diagrams are important in producing the bound states. The attraction from the direct diagram is enough to produce the $D_{12}$ state while no bound states can be produced with the direct diagram for the $D_{21}$ and $D_{22}$ states. These two states are mainly produced from the contributions from the cross diagram.  Such a result suggests that the cross diagram is important and can not be neglected in the calculation.

\section{Summary and discussion}\label{Sec: summary}

Inspired by the experimental hint of the dibaryon $N\Delta(D_{21})$ at WASA-at-COSY, we study the possible molecular states from the $N\Delta$ interaction. Within the one-boson-exchange model, the interaction is constructed with the help of the effective Lagrangians, whosw coupling constants are determined by experiment and SU(3) symmetry. After inserting the potential  into the qBSE, we search for the bound states from the $S$-wave $N\Delta$ interaction.

Among four states considered in the current work, three  bound states, $D_{12}$, $D_{21}$, and $D_{22}$, can be found in the range of the parameter $\alpha$ considered here. We also perform a crude calculation about the deuteron within the current theoretical frame for reference. The deuteron can be reproduced from the $NN$ interaction with a parameter a little smaller than the one for $D_{21}$ state determined by the experimental mass.   The results suggest that the $\pi$ exchange plays the most important role in producing these bound states. The $\rho$ exchange provides a marginal contribution to produce the $D_{12}$ state while the $\omega$ exchange will weaken the interaction. The binding of the  $D_{21}$ state is deepest among the three states. With values of the parameter $\alpha$ for which the experimental value of the binding energy for $D_{21}$ state is obtained, the other two states are predicted with much smaller binding energy.

Here, we would like to address the possible uncertainties in the current work. In our models, the spectator approximation and the replacement of $q^2$ by $-|q^2|$ in the propagator for the cross diagram will introduce model uncertainties. Such uncertainties will be absorbed by the parameter $\alpha$. Hence, $\alpha$ can vary a little. Besides, as discussed in the above section. The deuteron is reproduced at $\alpha=2.7$ which is smaller than the value of 3.2 suggested by the experimental mass of $N\Delta(D_{21})$ at WASA-at-COSY. With such a value, the binding energy of $D_{22}$ state becomes very  small, and the $D_{12}$ even disappears. Considering the model uncertainties, these two states, and  have very small binding energies, and may even not exist. 

The $d^*(2380)$ is the second observed dibaryon besides the deuteron. However, it seems to be a compact hexaquark instead of a molecular state like the deuteron. It is interesting to find more dibaryons to understand the internal structure of the dibaryons. The masses of the dibaryon $N\Delta(D_{21})$ suggested by the  WASA-at-COSY Collaboration is close  the $N\Delta$ threshold, and it has a width very close to the sum of widths of a nucleon and a $\Delta$ baryon, which supports it as a molecular state.  However, the experimental hint of the state $N\Delta(D_{21})$ at WASA-at-COSY is very weak and is not confirmed by other experiments. The existence of such state requires further theoretical and experimental studies. Based on our work, the existence of $N\Delta(D_{21})$ suggests the possible existence of other two $N\Delta$ molecular states, $D_{21}$ and $D_{22}$. It is interesting to search for such states in the experiment.

\section*{Acknowledgments}

This project is  supported by the National
Natural Science Foundation of China (Grant No.11675228).

\end{document}